\documentclass[showpacs,amsmath,amssym,twocolumn,notitlepage]{revtex4-1}

\usepackage{graphicx}% Include figure files
\usepackage{dcolumn}% Align table columns on decimal point
\usepackage{bm}% bold math

 \let\b=\beta       \let\d=\delta  
         
\let\m=\mu                      \let\r=\rho
\let\s=\sigma         
   \let\o=\omega 
      \let\L=\Lambda

\def\\{\hfill\break} \let\==\equiv

\let\io=\infty 

\let\0=\noindent 

\def\media#1{{\langle#1\rangle}}
\def\eg{\hbox{\it e.g.\ }}

\def\tende#1{\,\vtop{\ialign{##\crcr\rightarrowfill\crcr
 \noalign{\kern-1pt\nointerlineskip}
 \hskip3.pt${\scriptstyle #1}$\hskip3.pt\crcr}}\,}
\def\otto{\,{\kern-1.truept\leftarrow\kern-5.truept\to\kern-1.truept}\,}

\def\VV{{\cal V}}

\def\BBB{{\cal B}}

\def\T#1{{#1_{\kern-3pt\lower7pt\hbox{$\widetilde{}$}}\kern3pt}}
\def\VVV#1{{\underline #1}_{\kern-3pt
\lower7pt\hbox{$\widetilde{}$}}\kern3pt\,}
\def\W#1{#1_{\kern-3pt\lower7.5pt\hbox{$\widetilde{}$}}\kern2pt\,}
\def\Tr{\rm Tr\,}

\def\indica{\leaders \hbox to 0.5cm{\hss.\hss}\hfill}
\def\guida{\leaders\hbox to 1em{\hss.\hss}\hfill}
\mathchardef\oo= "0521

\def\V#1{{\bf #1}}\def\xx{{\bf x}}\def\yy{{\bf y}}\def\kk{{\bf k}}
\def\pp{{\bf p}}
\def\bT{{\bf T}}

\def\be{\begin{equation}}
\def\ee{\end{equation}}
\def\bea{\begin{eqnarray}}\def\eea{\end{eqnarray}}
\def\bp{\begin{pmatrix}}
\def\ep{\end{pmatrix}}

\def\qed{\raise1pt\hbox{\vrule height5pt width5pt depth0pt}}

\begin{document}

%\preprint{APS/123-QED}

\title{Rigorous construction of ground state
correlations in graphene:\\
renormalization of the velocities and Ward Identities}

\author{Alessandro Giuliani}
%\email{giuliani@mat.uniroma3.it}
\affiliation{%
Universit\`a di Roma Tre,
L.go S. L. Murialdo 1, 00146 Roma, Italy}
\author{Vieri Mastropietro}%
% \email{mastropi@axp.mat.uniroma2.it}
\affiliation{%
Universit\`a di Roma Tor Vergata,
Viale della Ricerca Scientifica 00133 Roma, Italy}

\begin{abstract} We consider the two-dimensional Hubbard model on the honeycomb lattice,
as a model for single layer graphene with screened Coulomb
interactions; at half filling and weak coupling, we construct its
ground state correlations by a convergent multiscale expansion,
rigorously excluding the presence of magnetic or superconducting
instabilities or the formation of a mass gap. The
Fermi velocity, which can be written in terms of a
convergent series expansion, remains close to its non-interacting value
and turns out to be isotropic; as a consequence, the Dirac cones 
are isotropic at low energies. On the contrary, the interaction produces an asymmetry
between the two components of the charge velocity, in contrast with the
predictions based on
relativistic or continuum approximations.
\end{abstract}


\maketitle

The recent experimental realization of a monoatomic graphitic film
\cite{N}, known as {\it graphene}, has elicited an enormous
interest in the study of the properties of two-dimensional
electron systems on the honeycomb lattice, which is the typical
underlying structure displayed by single--layer graphene sheets.
Graphene is quite different from most conventional quasi--two
dimensional electron gases, because of the peculiar
quasi--particles dispersion relation, which closely resembles the
one of massless Dirac fermions in $2+1$ dimensions \cite{W}.
Already in the absence of interactions, the system displays highly
unusual features, such as the anomalous dependence of the
cyclotron mass on the electronic density, an anomalous integer
quantum Hall effect, and the insensitivity to localization effects
generated by disorder. In real systems, despite the unavoidable
presence of electron-electron interactions, such consequences of
the relativistic-like dispersion relation have been experimentally
verified \cite{G,G1}; the observation based on angle-resolved
photoemission spectroscopy \cite{Z,B1} are compatible with linear
dispersion relation and isotropic Fermi velocity.

A basic model for investigating the effect of the electron-electron
interactions in graphene is the 2D Hubbard model on the honeycomb
lattice, in the presence either of a short or of a long range
interaction, corresponding to the cases of screened or unscreened
Coulomb interactions, respectively; see
\cite{S,JP,V1,V2,V3,V4,Mis,SS,KUC, HJV}. Usually the analysis
of this model is performed by mean field and Renormalization Group
(RG) methods at the lowest perturbative orders
\cite{S,JP,V1,V2,V3,V4,Mis,SS,KUC,HJV}, neglecting the
presence of the lattice and replacing the exact dispersion
relation by its linear approximation around the Dirac points.
However, in such analyses there is no control of the effects
produced by the truncations of the exact RG equations or by the
considered approximations, so that non-perturbative effects, such
as a mass generation, have not been excluded so far and are still
subject of an active debate \cite{D,GSC,JR,H06,K08}. Moreover,
possible symmetry-breaking effects due to the presence of the
underlying lattice have not been considered so far, and it remains
to be seen whether the fact that Lorentz invariance is explicitly
broken by the lattice can induce anisotropic renormalizations of
the Dirac cones, or of the spin and charge velocities.

It would be desirable to substantiate the predictions of
theoretical analysis by rigorous results and exact solutions and,
in the case of controverse issues, to be able to rigorously
exclude one conclusion or the other. Unfortunately, there are very
few rigorous results about the structure of the ground state of
the Hubbard model in two or more dimensions, among which we would
like to mention the results in \cite{Lieb}, guaranteeing the
uniqueness of the ground state and the vanishing of its total
spin. However, as far as we know, so far no results were proved
about existence or non existence of long range order and about the
long distance behavior of correlation in the Hubbard model on the
honeycomb or other lattices.

Two theorems reporting the first rigorous construction of the
ground state correlations in the 2D Hubbard model on the honeycomb
lattice at half filling, weak coupling and short range interactions
are stated here. Our results
exclude the presence of magnetic or superconducting instabilities
and the formation of a mass gap. The interaction changes by a finite amount
the wave function renormalization and the Fermi velocity.
Note that the interacting Fermi velocity remains
isotropic, even though the model breaks the invariance under $90^o$ degree
rotations; the isotropy of the Fermi velocity implies
the isotropy of the Dirac cones at low energies.

On the contrary, we predict that the charge velocity develops an
asymmetry between the two components, an effect that is in principle accessible
to experiments. Note that the latter conclusion is in contrast with the
naive expectation that weak short range interactions, being irrelevant in the
RG sense, do not alter the Dirac spectrum and the spin and charge
velocities. In the case of unscreened Coulomb interactions, we are not yet able
to control the convergence of the renormalized series; however, even
in this case, we predict that the Fermi velocity remains isotropic at all
orders, while the isotropy of the charge velocity is broken already at second
order, as in the case of short ranged interactions.

Our analysis is based on the methods of {\it constructive
renormalization}, which have already proven to be quite effective
in analyzing 1D interacting fermionic systems in their ground state
\cite{BM} and 2D systems up to exponentially small temperatures or in the
presence of a non-symmetric Fermi surfaces \cite{R,BGM,FKT}.
While constructive renormalization (see \eg \cite{M}
for an introduction)
is based on the RG ideas,
the way in which it implements these ideas is slightly different
with respect to other more standard schemes, the main advantages
being that the resulting method is: (i) {\it exact},
in the sense that it does not need any relativistic approximation or continuum
limit and it allow us to keep the full lattice structure of the problem;
(ii) {\it non-perturbative}, in the sense that it involves expansions whose
convergence can be mathematically proved.

The Hamiltonian of the 2D Hubbard model on the honeycomb lattice
at half filling in second quantized form is given by:
\bea&& H_\L=-\sum_{\substack{\vec x\in \L \\
i=1,2,3}}\sum_{\s=\uparrow\downarrow} \Big( a^+_{\vec x,\s} b_{\vec
x+\vec \d_i,\s}^-+ b_{\vec x+\vec \d_i,\s}^+ a^-_{\vec x,\s} \Big)
+\label{1.1}\\
&&+ U\sum_{\substack{\vec x\in \L\\
i=1,2,3}}\sum_{\s,\s'}\Big(a^+_{\vec x,\s}
a^-_{\vec x,\s}-\frac12\Big)\Big(b_{\vec x+\vec \d_i,\s'}^+b_{\vec
x+\vec \d_i, \s'}^--\frac12\Big)\;,\nonumber\eea
where $\L$ is a periodic triangular lattice with basis ${\vec
a_{1,2}}=\frac12(\pm\sqrt{3},3)$
and the nearest neighbor vectors
$\vec \d_i$ are defined as ${\vec \d_1}=(0,1)$, ${\vec
\d_{2,3}}=\frac12 (\pm\sqrt{3},-1)$.
The creation and annihilation fermionic
operators with spin index $\s=\uparrow\downarrow$, $a^\pm_{\vec
x,\s}, b^\pm_{\vec x+\vec\d_i, \s}$, satisfy periodic boundary
conditions in $\vec x$. The choice of the interaction is done only
for definiteness (it is the simplest one for which the anisotropy of the charge
velocity is visible at first order in renormalized perturbation theory)
but our results are valid for a generic short range density-density
interaction.

We introduce the two component fermionic operators $\psi^\pm_{\vec
x,\s}= \big(a^\pm_{\vec x,\s}, b^\pm_{\vec x+\vec\d_1, \s}\big)$
and $\psi^\pm_{\xx,\s}= e^{H_\L x_0}\psi^\pm_{\vec x,\s} e^{-H_\L
x_0}$ with $\xx=(x_0,\vec x)$. If
$\media{\cdot}=\lim_{\b,|\L|\to\io} \Tr\{e^{-\b
H_\L}\bT\{\cdot\}\}/\Tr\{e^{-\b H_\L}\}$, with $\bT$ the fermionic
time ordering operator, the zero temperature $2n$-point Schwinger
functions are defined as
$\media{\prod_{i=1}^n\psi^{-}_{\xx_i,\s_i}
\psi^{+}_{\yy_i,\s'_i}}$. In the non interacting $U=0$ case, the
Fourier transform of the 2-point Schwinger function is given by
\be \hat
S_0(\kk)=\media{\hat\psi_{\kk,\s}^-\hat\psi_{\kk,\s}^+}\big|_{U=0}=
\bp -i k_0 && -v^*(\vec k)\cr -v(\vec k)&& -ik_0
\ep^{\!\!\!-1}\;,\label{prop}\ee
with $\hat\psi^\pm_{\kk,\s}=\int_{-\b/2}^{\b/2}dx_0\sum_{\vec
x\in\L} e^{\mp i\kk\xx}\psi^\pm_{\xx,\s}$ and $v(\vec
k)=\sum_{i=1}^3e^{i\vec k(\vec\d_i-\vec\d_1)}$. $\hat S_0(\kk)$ is
singular at the {\it Fermi points}
$\pp_F^{\pm}=(0,\vec p_{F}^{\ \pm})$, where $\vec
p_{F}^{\ \pm}=(\pm{\frac{2\pi}{3\sqrt{3}}},\frac{2\pi}{3})$.
Close to $\vec p_F^{\,\pm}$, $v(\vec k'+\vec
p_F^{\,\pm})\simeq (3/2)(\pm k_1'+ik_2')$, so that the free Schwinger function is
asymptotically the same as the one of massless Dirac fermions in
$2+1$ dimension.

The {\it density} operator is defined as
$\hat\rho_\pp=(\b|\L|)^{-1}$ $\sum_{\kk,\s}\hat\psi^+_{\kk,\s}
\hat\psi^-_{\kk-\pp,\s}$ and the definition of the current, for $U=0$,
is obtained from the equation $d\r_{\xx}/dx_0=[H_\L,\r_{\xx}]$: in fact,
the latter, for small $\pp$, assumes the form of a continuity equation
provided that the current is chosen as ${\hat\jmath}_{\pp,i}=
(\b|\L|)^{-1} \sum_{\kk,\s}
\hat\psi^+_{\kk,\s}\s_i\hat\psi^-_{\kk-\pp,\s}$, $i=1,2$, with
$\s_1,\s_2$ the first two Pauli matrices. The continuity
equation implies the validity of an asymptotic Ward Identity:
defining $\kk'=\kk-\pp_F^\pm$, if $\kk',\pp,\kk'-\pp$ are small and
of the same order of magnitude \cite{.pdf},
\bea &&\media{\big(ip_0\hat\r_\pp\pm (3/2)p_1 \hat \jmath_{\pp,1}+(3/2) p_2 \hat
\jmath_{\pp,2}
\big); \hat\psi^{-}_{\kk,\s}\hat\psi^{+}_{\kk-\pp,\s}}\simeq
\nonumber\\
&&\hskip1.truecm
\simeq [\media{\hat\psi^{-}_{\kk,\s}\hat\psi^{+}_{\kk,\s}}
-\media{\hat\psi^{-}_{\kk-\pp,\s}\hat\psi^{+}_{\kk-\pp,\s}}]\;.
\label{wi}\eea

When the interaction is present, the Schwinger functions are not
exactly computable anymore; however, quite remarkably, they can be
computed in terms of a {\it convergent} renormalized perturbative
series, and their long distance asymptotic properties can be
rigorously derived, as summarized by the following theorem.
\vskip.4cm
{\it THEOREM 1: There exists a constant $U_0>0$ such that, if $|U|\le U_0$,
the specific ground state energy and the
zero temperature Schwinger functions of model (\ref{1.1}) are
analytic functions of $U$. The Fourier transform of the 2-point
Schwinger function $S(\kk)$ is singular only at $\kk=\pp_F^{\pm}$ and,
close to the singularity, it can be written as \cite{.pdf},
\be S(\kk)\simeq\frac{1}{Z}\bp -i k_0 & v_F(\mp k_1'+i k_2') \cr v_F(\mp
k_1'-i k_2') & -ik_0\ep^{\!\!-1}\;,
\label{thm1}\ee
where $Z=Z(U)$ and $v_F=v_F(U)$
are analytic functions of $U$, such that $Z=1+O(U^2)$ and
$v_F=3/2+bU+O(U^2)$, with
\be b=\int_{B_1}\frac{d\vec k}{2|B_1|} \frac{v(\vec k)}{|v(\vec
k)|}\partial_{p_1} v(\vec p-\vec k)\Biggr|_{\vec
p=\vec p_F^{\,+}}=0.511\ldots\label{aa}\ee
and $B_1$ the first Brillouin zone.}

The result summarized in Theorem 1 says that the interaction does
not qualitatively change the asymptotic behavior of the 2-point Schwinger
function close to the Fermi points; the effect of the interaction
is essentially to change by a finite amount the wave function
renormalization and the Fermi velocity. This implies that the
interacting correlations decay as fast as in the non-interacting
case and, therefore, the presence of quantum instabilities in the
ground state, such as N\'eel or superconducting long range order,
is rigorously excluded at half filling and weak
coupling, together with the possibility of a mass generation.

Note also that, in the presence of the interaction, the Fermi velocity
remains the same in the two coordinate directions even though the
model does not display $90^o$ discrete rotational symmetry, but
rather a $120^o$ rotational symmetry; such a remarkable property
can be easily checked at first order (replacing
$\partial_{p_1}$ by $-i\partial_{p_2}$ in Eq.(\ref{aa}),
the same result is found); for
a proof at all orders in the convergent expansion for $v_F$, see
below. The isotropy of the Fermi velocity implies
the isotropy of the Dirac cones at low energies.
\vskip.3cm

{\it THEOREM 2: For $|U|\le U_0$, if $\kk',\pp,\kk'-\pp$ are small and
of the same order of magnitude \cite{.pdf}, then
\begin{eqnarray}
&&\media{\big(ip_0\hat\r_\pp\pm  v_1 p_1 \hat\jmath_{\pp,1}+ v_2
p_2 \hat\jmath_{\pp,2}\big);
\hat\psi^{-}_{\kk,\s}\hat\psi^{+}_{\kk-\pp,\s}}
\simeq\nonumber\\
&&\hskip.8truecm \simeq
[\media{\hat\psi^{-}_{\kk,\s}\psi^{+}_{\kk,\s}}
-\media{\hat\psi^{-}_{\kk-\pp,\s}\hat\psi^{+}_{\kk-\pp,\s}}]\;,
\label{thm2}\end{eqnarray}
where the charge velocity $v_{1,2}=3/2+O(U)$ is analytic in $U$
and $v_1-v_2=a U+O(U^2)$, with
\be\label{xx} a=\frac{3}{4}\int_{B_1}\frac{d\vec k}{|B_1|}\frac{v^2(\vec k)}{
|v(\vec k)|^3} v^*(\vec k-\vec p_F^{\,+})=-0.03165\ldots\ee
}

Theorem 2 says that, in the presence of interactions, a new Ward
Identity, {\it different} from the non-interacting one, is
verified, the main difference with respect to Eq.(\ref{wi}) being
that the charge velocity $(v_1,v_2)$ is interaction-dependent and
different from the Fermi velocity. Remarkably, its two components
are different: this anisotropy is related to the presence of the
lattice, that is to the irrelevant terms in a RG sense, which are
not negleted in our exact scheme. In fact, if we replace $v(\vec
k'+\vec p_F^{\,\pm})$ by its asymptotic expression $(3/2)(\pm
k_1'+i k'_2)$, the anisotropy coefficient defined by Eq.(\ref{xx})
vanishes exactly (and so do the higher order corrections).
We remark that, while
Theorems 1 and 2 refer to the case of short range interactions,
the conclusions concerning the symmetry of the Fermi and charge
velocities remain true, as statements at all orders, even for the
case of unscreened Coulomb interactions.

We now sketch the proof of the two theorems above (for a detailed
proof we refer to \cite{GM}). The starting point is the
well-known representation of the ground state energy in terms of a
Grassman functional integral: $e_0=\lim_{\b,|\L|\to\io}
(\b|\L|)^{-1}\log\int P(d\psi)\exp\{\VV(\psi)\}$, where $P(d\psi)$
is the Grassman gaussian integration with propagator $\hat
S_0(\kk)$, see Eq.(\ref{prop}), and $\VV(\psi)$ is the quartic
interaction Eq.(\ref{1.1}).
One can compute $e_0$ by expanding the exponential $\exp\{\VV(\psi)\}$
in Taylor series in $U$ and naively integrating term by term the Grassmann
monomials, using the Wick rule; however, by such procedure, it is very
difficult to take into account the cancellations present in the perturbative
series. The bounds obtained by this ``simple'' procedure are non-uniform in
$\b$, and they do not allow one to take the thermodynamic and zero temperature
limits. Therefore, we set up an iterative procedure for the computation of
$e_0$, based on (Wilsonian) Renormalization Group (RG) and involving non
trivial resummations of the perturbative series.

The first step is to decompose the
propagator $\hat S_0(\kk)$ as a sum of propagators supported close to
the two Fermi points and more and more singular
in the infrared region, labeled by a quasi particle index $\o=\pm$ (labelling
the Fermi points) and by an integer $h\le 0$,
so that $\hat S_0(\kk)=
\sum_{h\le 1}^{\o=\pm}\hat g_\o^{(h)}(\kk-\pp_F^\o)$,
with $\hat g_\o^{(h)}$ supported on quasi-momenta of scale $2^h$ and,
on the support, of size $||\hat g_\o^{(h)}||\sim 2^{-h}$.
At this point, we compute $e_0$ by iteratively integrating the propagators
$\hat g^{(0)},\hat g^{(-1)},\ldots$ After each integration step we rewrite
\be e_0=F_h+\lim_{\b,|\L|\to\io}\frac1{\b|\L|}
\log\int \prod_{\o=\pm}P(d\psi_\o^{(\le h)})
e^{\VV^{(h)}(\psi^{(\le h)})}\;,\ee
where $P(d\psi_\pm^{(\le h)})$ is the Grassmanian quadratic integration with
propagator given by
\be g_\pm^{(\le h)}(\kk')\simeq \frac{\chi_h(\kk')}{Z_h}\bp -ik_0 &
c_h(\mp k_1'+ik_2')
\cr c_h(\mp k_1'-ik_2') & -ik_0\ep^{\!\!\!-1}\ee
where $\chi_h^{-1}(\kk')$ is a
smooth compact support function nonvanishing only for $|\kk'|\le 2
^h$; $\VV^{(h)}$ is the {\it effective potential}, a sum of
monomials of arbitrary order, with kernels that are {\it analytic
functions} of $U$: analyticity is a very non trivial property
obtained exploiting anticommutativity properties of Grassmann
variables, via {\it Gram inequality} for determinants. The scaling
dimensions of the kernels with $n_e$ external lines are equal to
$3-n_e$, modulo an additional dimensional gain, following from the
fact that all kernels with $\ge 4$ external lines are irrelevant
in a RG sense (see \cite{GM}, Theorem 2). The kernels $\hat
W_2^{(h)}(\kk')$ with $n_e=2$, which are linearly relevant, can be
inserted step by step in the gaussian integration, thanks to the
fact that they have the same reality/symmetry properties as the
free quadratic action: in particular, it is found that $\hat
W^{(h)}_2(\V0)=0$ and
\be \kk'\partial_{\kk'} \hat W^{(h)}_2(\V0)=
\bp -i z_hk_0 & \d_h(\mp k_1'+i k_2') \cr \d_h(\mp k_1'-i k_2')
& -iz_hk_0\ep \;,\label{quad}\ee
for suitable real constants $z_h,\d_h$. Note that $\mp k_1'$ and
$ik_2'$ are multiplied by the same constant $\d_h$, which is quite
remarkable; see \cite{GM}, Lemma 2, for a proof. Iterating the
procedure above, we find recursive equations for $Z_h$ and
$c_h$; in the $h\to-\io$ limit, the two running coupling
constants converge to values $Z_{-\io}=Z=Z(U)$ and $c_{-\io}=v_F=v_F(U)$,
which are
close to their unperturbed values and are analytic in $U$ (again,
thanks to the fact that all sub-diagrams with $n_e\ge 4$ are
irrelevant in a RG sense). This completes the discussion
concerning analyticity of $e_0$. A similar discussion allows us to
prove analyticity of the Schwinger functions and Eq.(\ref{thm1}),
see \cite{GM}, Sec.III.D.

Let us now discuss a sketch of the proof of Theorem 2. We perform
a multiscale analysis similar to the one sketched above, with
$\VV(\psi)$ replaced by $\VV(\psi)-\BBB(\phi,J)$, with
$\BBB(\phi,J)=(\phi^+,\psi^-)+(\psi^+,\phi^-)+\sum_{\m=0}^2(J_\m,j_\m)$,
$\phi^\pm_{\xx,\s}$ two external Grassmann fields,
$J_{\xx,\m}$, an external commuting field, and $j_{\xx,\m}$ the
current (here $j_{\xx,0}=\rho_\xx$, with $\rho$ the density, see the lines preceding Eq.(\ref{wi})).

The iterative integration procedure described above, in this case,
produces, besides the effective potential $\VV^{(h)}$, new terms
involving the external fields. In particular, at scale $h$, the
effective source term is given by
$\sum_{\m,\m'}Z^{(h)}_{\m,\m'}(J_{\m},j_{\m'}^{(\le h)})$, with
$Z_{\m,\m'}=\d_{\m,\m'}Z_{\m}$, by the discrete invariance
symmetries of the model, see \cite{GM}, Lemma 1. A crucial
remark is that, while in a relativistic QFT $Z^{(h)}_{\m}$ is
$\m$-independent, here it is not, the relativistic symmetry being
broken by the presence of the underlying lattice (i.e., by the
irrelevant terms in the fermionic action). We find that, in the
limit $h\to-\io$, $Z^{(h)}_{\m}\to Z_\m=Z_\m(U)$, which are
analytic functions of $U$, with $Z_0=Z=1+O(U^2)$ and
$Z_2-Z_1=(2a/3)U+O(U^2)$, with $a$ given in Eq.(\ref{xx}). See Fig.1.

\begin{figure}[ht]
\centering
\includegraphics[width=.25\textwidth,angle=0]{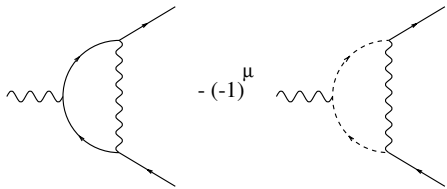}
\caption{The two graphs contributing at first order to $Z_{\m}$, $\m=1,2$.
The full line corresponds to a diagonal propagator and the dotted line to
an off-diagonal one. Note that
the two graphs appear with a different sign, depending on
the value of $\m$. The second graph is vanishing in the
continuum approximation $v(\vec k)\simeq (3/2)(\pm k_1'+ik_2')$.}
\end{figure}

Using the fact that all kernels with four or more external lines are
irrelevant, we find that, for $\kk',\kk'-\pp,\pp$ small,
\be\media{\hat\jmath_{\pp,\m};\hat\psi^-_{\kk-\pp,\s}\psi^+_{\kk,\s}} \simeq
Z_\m S(\kk)\s_\m S(\kk-\pp)\;,\ \m=0,1,2\label{wi2}\ee
with $\s_0=\openone$ and $S(\kk)$ the interacting 2-point Schwinger function in
Eq.(\ref{thm1}). The
combination $S(\kk)\s_\m S(\kk+\pp)$ is asymptotically the same
as the vertex of a relativistic QFT with wave
function renormalization $Z$ and ``speed of light'' equal to
$v_F$; therefore, it satisfies a relativistic WI, relating it to
the derivative of $S(\kk)$:
\bea && S(\kk)\left[ip_0+v_F(\pm p_1\s_1+p_2\s_2)\right]
S(\kk-\pp) \simeq\nonumber\\
&&\hskip1.truecm \simeq\frac1{Z}\big[S(\kk)-S(\kk-\pp)\big]\;.\label{wi3}\eea
Combining Eqs.(\ref{wi2}) and (\ref{wi3}), and recalling that,
by symmetry, $Z=Z_0$,
we get Eq.(\ref{thm2}) with $v_{1,2}=v_F Z_0/Z_{1,2}$.

In conclusion, we analyzed the properties
of single layer graphene at half filling, 
described by a Hubbard model on the honeycomb lattice, 
going for the first time
beyond the approaches based on finite-order truncations and relativistic 
approximations (previous analysis  
taking into account lattice effects were focusing on doped graphene, see
\eg \cite{V3,V4a,V4b,V5}). 
In the case of short range
interactions, we proved analyticity of the theory at weak coupling; this
gives a rigorous confirmation to the belief, see \eg \cite{JR}, that
non-perturbative effects such as 
quantum instabilities or the opening of a mass gap can be possibly present only
at large values of the coupling. 
We proved that the Fermi velocity and the Dirac cones at low energies are isotropic, in
agreement with observations based on angle-resolved photoemission
spectroscopy \cite{Z,B1}; previous analyses based on
relativistic approximations were inconclusive in this respect, since the
(previously neglected) symmetry-breaking terms due to the lattice 
produce a renormalization of the Fermi velocity, which could in principle be
anisotropic. 
This is by no means just an academic possibility;
indeed, while this anisotropy effect is not visible in the Fermi
velocity, we show that it is observable in other quantities, like the charge
velocity appearing in the Ward Identities, which turns out to be asymmetric 
in the two coordinate directions.
This asymmetry was previously unnoticed and it may be detected in
future experiments. Finally, we stress that the assumption 
of local or short range interaction plays a crucial role in our analysis;
the unscreened Coulomb interactions is {\it marginal}
instead of {\it irrelevant} in the RG sense, and its effect on the 
physical properties can be in principle much more relevant.
The unscreened Coulomb interactions has been studied
up to now mainly in the relativistic approximation and
at lowest perturbative orders, starting from \cite{V1,V2}, and we believe that going beyond 
such approximations will give a definite answer to the
question of the role of the interactions in the properties of real graphene.
\vskip.4cm
We thank G. Benfatto for many useful discussions.


\begin{thebibliography}{9}
\bibitem{N} K. S. Novoselov et al.,
{\it Science} {\bf 306}, 666 (2004).

\bibitem{W} P. R. Wallace, {\it Phys. Rev.} {\bf 71}, 622 (1947).

\bibitem{G} K. S. Novoselov et al., {\it Nature (London)}
{\bf 438}, 197 (2005).

\bibitem{G1} A. K. Geim and K. S. Novoselov, {\it
Nature Mater.} {\bf 6}, 183 (2007).

\bibitem{Z} S. Y. Zhou et al., {\it Nat. Phys.} {\bf 2}, 595 (2006).

\bibitem{B1} A. Bostwick et al., {\it Nat. Phys.} {\bf 3}, 36 (2007).

\bibitem{S} G. W. Semenoff, {\it Phys. Rev. Lett.} {\bf 53}, 2449 (1984).

\bibitem{JP} R. Jackiw and S.-Y. Pi, {\it  Phys. Rev. Lett.} {\bf 98}, 266402 (2007).

\bibitem{V1} J. Gonzalez et al., 
{\it  Nucl. Phys. B} {\bf 424}, 595 (1994).

\bibitem{V2} J. Gonzalez et al., {\it Phys. Rev. B}
{\bf 63}, 134421 (2001).

\bibitem{V3} S. Das Sarma et al.,  {\it
Phys. Rev. B} {\bf 75}, 121406(R) (2007).

\bibitem{V4} C.-H. Park et al., {\it Phys. Rev.
Lett.} {\bf 99}, 086804 (2007).

\bibitem{Mis} E. G. Mishchenko, {\it Phys. Rev. Lett} {\bf 98}, 216801 (2007).

\bibitem{SS} D. E. Sheehy, J. Schmalian, {\it Phys. Rev. Lett.} {\bf 99}, 226803 (2007).

\bibitem{KUC} V. N. Kotov et al., {\it Phys. Rev. B} {\bf 78}, 035119 (2008).

\bibitem{HJV} I. F. Herbut et al., {\it Phys. Rev. Lett.} {\bf 100}, 046403 (2008).

\bibitem{D} A. H. Castro Neto et al.,
{\it Rev. Mod. Phys.} {\bf 81}, 109 (2009).

\bibitem{GSC} V. P. Gusynin et al., {\it Int. J. Mod. Phys. B} {\bf 21},
4611 (2007).

\bibitem{JR} I. Herbut et al., {\it Phys. Rev. B} {\bf 79}, 085116 (2009).

\bibitem{H06} I. F. Herbut {\it Phys. Rev. Lett.} {\bf 97}, 146401 (2006).

\bibitem{K08} D. V. Khveshchenko, {\it J. Phys.: Condens. Matter} {\bf 21}, 075303 (2009).

\bibitem{Lieb} E. H. Lieb:
{\it Phys. Rev. Lett.} {\bf 62},
1201 (1989).

\bibitem{BM} G. Benfatto and V. Mastropietro, {\it Comm. Math. Phys.}
{\bf 258}, 609 (2005).

\bibitem{R}
M. Disertori and V. Rivasseau, {\it Phys. Rev. Lett.} {\bf 85}, 361
(2000).

\bibitem{BGM} G. Benfatto et al., {\it Ann. Henri
Poincar\'e} {\bf 7}, 809 (2006).

\bibitem{FKT} J. Feldman et al., {\it Comm. Math. Phys.} {\bf 247},
1 (2004).

\bibitem{M} V. Mastropietro, {\it Non-perturbative Renormalization} (World Scientific, Singapore, 2008).

\bibitem{.pdf} The simbol ``$\simeq$'' means
that the ratio of the l.h.s. to the r.h.s.
is equal to $1+R$, with $R$ a
matrix computed in terms of a convergent series and satisfying
$||R||\le C_\theta|\kk-\pp_F^\pm|^\theta$ for any $0<\theta<1$ and a
suitable constant $C_\theta$.

\bibitem{GM} A. Giuliani and V. Mastropietro, {\it Comm. Math. Phys.} {\bf 293}, 30 (2010).

\bibitem{V4a} R. Roldan et al., {\it Phys. Rev. B} {\bf 77}, 115410 (2008)

\bibitem{V4b}  B. Valenzuela and M. Vozmediano, {\it New J. Phys.} {\bf 10}, 113009 (2008).

\bibitem{V5} M. Polini et al., {\it Solid State Commun.} {\bf 143}, 58 (2007).

\end{thebibliography}
\end{document}

%\preprint{APS/123-QED}

\title{Erratum: Rigorous construction of ground state
correlations in graphene:
renormalization of the velocities and Ward Identities\\ $[$Phys. Rev. B
79, 201403(R) (2009)$]$}

\author{Alessandro Giuliani and Vieri Mastropietro}

\pacs{05.10Cc, 05.30.Fk, 71.10.-w}
\maketitle

\setcounter{equation}{2}

While the results stated in Theorem 2 of \cite{GM} are mathematically
correct, their physical interpretation is not. The bilinear
operators $\r$ and $\vec\jmath$ defined in \cite{GM} do not correspond to the 
physical (Berry-gauge invariant) lattice density and lattice current operators \cite{Hal}; 
therefore, the anisotropy in the renormalization constants of $\r$ and 
$\vec\jmath$ does not imply any physical anisotropy in the charge
velocity, contrary to what is claimed in \cite{GM}. Even so, 
the anisotropy found in \cite{GM} has some interest: it illustrates the 
fact that irrelevant terms can produce observable effects, e.g., they can 
produce
anisotropies that would be absent if one just neglected all the lattice 
effects from the beginning.
On the other hand, the multiscale analysis described in \cite{GM} 
can be also used to analyze the renormalization of the correct lattice density and current
operators. The result of Theorem 2 and its interpretation should be corrected as follows. 
\begin{enumerate}
\item The last sentence of the abstract should be changed into: ``Ward Identities 
imply exact relations between the vertex functions, the wave function renormalization and
the Fermi velocity."
\item At p.1, the fifth paragraph should be completely eliminated (from ``On the contrary..." to 
``... short-ranged interactions.")
\item At p.2, the fourth paragraph (from ``The {\it density} operator..." to Eq.(3)) 
should be changed into: ``The {\it density} operator is
defined as $\hat\rho_\pp=(\b|\L|)^{-1}$
$\sum_{\kk,\s}\hat\psi^+_{\kk,\s} \G_0(\vec k,\vec p)\hat\psi^-_{\kk-\pp,\s}$, with 
$[\G_0(\vec k,\vec p)]_{ij}=\d_{i,j}(\d_{i,1}+\d_{i,2} e^{-i\vec p\vec\d_1})$, and the
definition of the current is obtained from the equation
$d\r_{\xx}/dx_0=[H_\L,\r_{\xx}]$, which has the form of a continuity equation
$ip_0\hat\r_\pp+i\vec p\vec\jmath_\pp=0$, provided 
that the current is chosen as $\hat\jmath_{\pp,i}=(\b|\L|)^{-1}(3/2)\sum_{\kk,\s}\hat\psi^+_{\kk,\s} 
\G_i(\vec k,\vec p)\hat\psi^-_{\kk-\pp,\s}$, where $i=1,2$, with  
$$\vec\G(\vec k,\vec p)=\frac23\sum_{j=1}^3\vec\d_j
\h^j_{\vec p} \begin{pmatrix} 0& -e^{-i(\vec k-\vec p)(\vec\d_j-\vec\d_1)}\\
+e^{i\vec k(\vec\d_j-\vec\d_1)}&0
\end{pmatrix}$$
and $\h^j_{\vec p}=\frac{1-e^{-i\vec p\vec\d_j}}{i\vec p\vec\d_j}$. The continuity 
equation implies the validity of an exact lattice Ward identity (WI), valid both in the absence and in the presence of the interaction, which reads
\be \media{(ip_0\hat\r_\pp+i\vec p\cdot{\vec\jmath}_\pp);\hat\psi^-_{\kk,\s}\hat\psi^+_{\kk-\pp,\s}}
=\media{\hat\psi^+_{\kk,\s}\hat\psi^-_{\kk,\s}}-\media{\hat\psi^+_{\kk-\pp,\s}\hat\psi^-_{\kk-\pp,\s}}
\;,\label{3}\ee
where $\media{\cdot;\cdot}$ denotes truncated expectation."
\item At p.2-3, Theorem 2 should be replaced by: ``THEOREM 2. {\it 
For $|U|\le U_0$, if 
$\kk'=\kk-\pp_F^\pm,\pp, \kk'-\pp$ are small and of the same order of magnitude,$^{28}$
then\setcounter{equation}{5}
\begin{eqnarray}
&&\media{\hat\r_\pp;\hat\psi^{-}_{\kk,\s}\hat\psi^{+}_{\kk-\pp,\s}}\simeq
Z_0\media{\hat\psi^{-}_{\kk,\s}\psi^{+}_{\kk,\s}}
\media{\hat\psi^{-}_{\kk-\pp,\s}\psi^{+}_{\kk-\pp,\s}}\;,\nn\\
&&\media{\hat\jmath_{\pp,i};\hat\psi^{-}_{\kk,\s}\hat\psi^{+}_{\kk-\pp,\s}}\simeq
Z_i\media{\hat\psi^{-}_{\kk,\s}\psi^{+}_{\kk,\s}}\G_i(\vec p_F^\pm,\vec 0)\media{\hat\psi^{-}_{\kk-\pp,\s}\psi^{+}_{\kk-\pp,\s}}\;,
\end{eqnarray}
where $Z_0, Z_1, Z_2$ are analytic in $U$, with 
\be Z_0=Z\;,\qquad Z_1=Z_2=v_F Z\;.\ee
"} 
\item At p.3, the first paragraph (from ``Theorem 2 says..." to ``...Coulomb interactions.")
should be changed into: ``Theorem 2 says that the density vertex function $Z_0$ is
equal to the wave function renormalization $Z$, and that the current
vertex function $Z_1$ divided by $Z$ is equal to the Fermi
velocity; note also that the current vertex function is isotropic, that is, $Z_i$ is 
independent of the direction $i=1,2$."
\item At p.3, in the fifth paragraph, the sentence ``is given by $\sum_{\m,\m'}Z_{\m,\m'}^{(h)}
(J_\m,j_{\m'}^{(\le h)})$, with $Z_{\m,\m'}=\d_{\m,\m'}Z_\m$" should be changed into 
``is given by  $\sum_{\m,\m'}Z_{\m,\m'}^{(h)}
(J_\m,\widetilde\jmath_{\m'}^{\;(\le h)})$, with $\widetilde\jmath_{\pp,\m}^{\;(\le h)}=(\b|\L|)^{-1}\sum_{\kk,\s}
\hat\psi^{(\le h)+}_{\kk,\s}\G_\m(\vec k,\vec p)\hat\psi^{(\le h)-}_{\kk-\pp,\s}$ and 
$Z_{\m,\m'}^{(h)}=\d_{\m,\m'}Z_\m^{(h)}$"
\item At p.3, the last part of the fifth paragraph (from ``A crucial remark..." to ``See Fig.1.")
should be replaced by ``We find that in the limit
$h\to-\io$, $Z^{(h)}_{\m}\to Z_\m=Z_\m(U)$, which are analytic
functions of $U$."
\item At p.3, in the right hand side of Eq.(11), ``$\s_\m$" should be replaced by 
``$\G_\m(\vec p_F^\pm,\vec 0)$". In the following line, the words ``$\s_0=1$ and" should be eliminated. One more line below, the sentence from ``The combination..." to ``...with $v_{1,2}=v_F
Z_0/Z_{1,2}$." (see l.2 after Eq.(12)) should be changed into: ``Using the WI Eq.(3) we find that $Z_0=Z$ and $Z_1=Z_2=v_F Z$."
\item Fig.1 should be eliminated.
\item At p.4, second column, the sentence from ``This is by no means..." to ``...in future experiments." should be eliminated.
\end{enumerate}

We thank Prof. F. D. M. Haldane for pointing out the correct physical 
definition of the lattice current.